\documentclass[conference]{IEEEtran}
\usepackage{amssymb,amsfonts,amsmath}
\usepackage{dsfont}
\usepackage{balance}
\usepackage{cite}

\IEEEoverridecommandlockouts

\ifCLASSINFOpdf
   \usepackage[pdftex]{graphicx}
   \DeclareGraphicsExtensions{.pdf,.jpeg,.png}
\else
\fi
\usepackage{fixltx2e}

\def\(({\left(}
\def\)){\right)}
\def\[[{\left[}
\def\]]{\right]}

\newtheorem{claim}{Claim}

\newcommand{\be}{\begin{equation}}
\newcommand{\ee}{\end{equation}}
\newcommand{\bea}{\begin{eqnarray}}
\newcommand{\eea}{\end{eqnarray}}

\renewcommand{\d}{\text {d}}

\def\ind{{\mathbf{1}}}
\def\B{{\rm B}}
\let\Ho\H
\def\H{{\rm H}}
\def\C{{\rm C}}
\def\I{{\rm I}}

\def\P{{\rm P}}
\def\argmax{{\rm argmax}}
\newcommand{\EE}{{\mathbf{E}}}
\newcommand{\PP}{{\mathbf{P}}}
\newcommand{\RR}{{\mathbb{R}}}
\newcommand{\BEAS}{\begin{eqnarray*}}
\newcommand{\EEAS}{\end{eqnarray*}}
\newcommand{\BEA}{\begin{eqnarray}}
\newcommand{\EEA}{\end{eqnarray}}

\begin{document}
%
% paper title
% can use linebreaks \\ within to get better formatting as desired
\title{Clustering from Sparse Pairwise Measurements}
\author{\IEEEauthorblockN{Alaa Saade}
\IEEEauthorblockA{Laboratoire de Physique Statistique\\
 \'Ecole Normale Sup\'erieure, 24 Rue Lhomond\\
Paris 75005}
\\
\IEEEauthorblockN{Marc Lelarge}
\IEEEauthorblockA{INRIA and \'Ecole Normale Sup\'erieure \\
Paris, France}

\and

\IEEEauthorblockN{Florent
  Krzakala}
\IEEEauthorblockA{
  Sorbonne Universit\'es, UPMC Univ. Paris 06\\
  Laboratoire de Physique Statistique, CNRS UMR 8550 \\
  \'Ecole Normale Sup\'erieure, 24 Rue Lhomond, Paris\\
  }
\\
  
\IEEEauthorblockN{Lenka Zdeborov\'{a}}
\IEEEauthorblockA{Institut de Physique Th\'{e}orique \\ CEA Saclay and
  CNRS, France.}
}

% conference papers do not typically use \thanks and this command
% is locked out in conference mode. If really needed, such as for
% the acknowledgment of grants, issue a \IEEEoverridecommandlockouts
% after \documentclass

% for over three affiliations, or if they all won't fit within the width
% of the page, use this alternative format:
% 
%\author{\IEEEauthorblockN{Michael Shell\IEEEauthorrefmark{1},
%Homer Simpson\IEEEauthorrefmark{2},
%James Kirk\IEEEauthorrefmark{3}, 
%Montgomery Scott\IEEEauthorrefmark{3} and
%Eldon Tyrell\IEEEauthorrefmark{4}}
%\IEEEauthorblockA{\IEEEauthorrefmark{1}School of Electrical and Computer Engineering\\
%Georgia Institute of Technology,
%Atlanta, Georgia 30332--0250\\ Email: see http://www.michaelshell.org/contact.html}
%\IEEEauthorblockA{\IEEEauthorrefmark{2}Twentieth Century Fox, Springfield, USA\\
%Email: homer@thesimpsons.com}
%\IEEEauthorblockA{\IEEEauthorrefmark{3}Starfleet Academy, San Francisco, California 96678-2391\\
%Telephone: (800) 555--1212, Fax: (888) 555--1212}
%\IEEEauthorblockA{\IEEEauthorrefmark{4}Tyrell Inc., 123 Replicant Street, Los Angeles, California 90210--4321}}

% use for special paper notices
%\IEEEspecialpapernotice{(Invited Paper)}

% make the title area
\maketitle

\begin{abstract}
  % \boldmath
  We consider the problem of grouping items into clusters based
  on few random pairwise comparisons between the items. We introduce
  three closely related algorithms for this task: a belief propagation
  algorithm approximating the Bayes optimal solution, and two spectral
  algorithms based on the non-backtracking and Bethe Hessian
  operators. For the case of two symmetric clusters, we conjecture that these algorithms are asymptotically optimal in that they detect
  the clusters as soon as it is information theoretically possible to
  do so. We substantiate this claim for one of the spectral approaches we introduce. 
  % As a by-product, we prove a conjecture on the detectability
  % transition of the labeled stochastic block model in a special case.
\end{abstract}
% IEEEtran.cls defaults to using nonbold math in the Abstract.
% This preserves the distinction between vectors and scalars. However,
% if the conference you are submitting to favors bold math in the abstract,
% then you can use LaTeX's standard command \boldmath at the very start
% of the abstract to achieve this. Many IEEE journals/conferences frown on
% math in the abstract anyway.

% no keywords

% For peer review papers, you can put extra information on the cover
% page as needed:
% \ifCLASSOPTIONpeerreview
% \begin{center} \bfseries EDICS Category: 3-BBND \end{center}
% \fi
%
% For peerreview papers, this IEEEtran command inserts a page break and
% creates the second title. It will be ignored for other modes.
\IEEEpeerreviewmaketitle

\section{Introduction}

\subsection{Problem and model}
\label{sec:problem_and_model}
Similarity-based clustering is a standard approach to label items in a dataset based on some measure of their resemblance. In general, given a dataset $\{x_i\}_{i\in[n]}\in \mathcal{X}^n$, and a symmetric measurement function $s : \mathcal{X}^2 \to \mathbb{R}$ quantifying the similarity between two items, the aim is to cluster the dataset from the knowledge of the pairwise measurements $s_{ij} := s(x_{i},x_{j})$, for $1\leq i < j \leq n$. This information is conveniently encoded in a similarity graph, which vertices represent items in the dataset, and the weighted edges carry the pairwise similarities. Typical choices for this similarity graph are the complete graph and the nearest neighbor graph (see e.g. \cite{Tutorial} for a discussion in the context of spectral clustering).    

Here however, we will not assume the measurement function $s$ to quantify the \emph{similarity} between items, but more generally ask that the measurements be \emph{typically different} depending on the cluster memberships of the items, in a way that will be made quantitative in the following. For instance, $s$ could be a distance in an Euclidean space or could take values in a set of colors (i.e. $s$ does not need to be real-valued). %.
Additionally, we will not assume knowledge of the measurements for all pairs of items in the dataset, but only for $O(n)$ of them chosen uniformly at random. 
Sampling is a well-known technique to speed up computations by reducing the number of non-zero entries \cite{achlioptas2001fast}. The main challenge is to choose the lowest possible sampling rate while still being able to detect the signal of interest. 
% in order to speed up computation by reducing the number of non-zero entries is a well-known technique \cite{achlioptas2001fast} and one of the main challenges is to choose the highest possible sampling rate in order to still detect signal. 
In this paper, we compute explicitly this fundamental limit for a simple probabilistic model and present three algorithms allowing partial recovery of the signal above this limit. Below the limit, in the case of two clusters, no algorithm can give an output positively correlated with the true clusters. Our three algorithms are respectively a belief propagation algorithm and two spectral algorithms based on the non-backtracking operator and the Bethe Hessian. Although these three algorithms are intimately related, so far, a sketch of rigorous analysis is available only for the spectral properties of the non-backtracking matrix. From a practical perspective however, belief propagation and the Bethe Hessian are much simpler to implement and show even better numerical performance.

To evaluate the performance of our proposed algorithms, we construct a
model with $n$ items in $k$ predefined clusters of same average size $n/k$, by assigning to each item $i \in [n]$ a cluster label $c_{i}\in[k]$ with uniform probability
$1/k$.
%We call $f_{a},\ a\in[k]$ the fraction of items belonging to cluster $a$.
We assume that the pairwise measurement between an item in cluster $a$ and
another item in cluster $b$ is a random variable with density
$p_{a,b}$.
% We generate the measurement graph as an Erd\Ho{o}s-R\'enyi random
% graph $G=(V,E)\in {\cal G}(n,\alpha/n)$. 
We choose the observed pairwise measurements uniformly at random, by generating an Erd\Ho{o}s-R\'enyi random graph $G=(V=[n],E)\in {\cal G}(n,\alpha/n)$. 
The average degree $\alpha$ corresponds to the sampling rate: pairwise measurements are observed only on the edges of $G$, and $\alpha$ therefore controls the difficulty of the problem. 
From
the base graph $G$, we build a measurement graph by %first assigning a cluster label $c_{i}\in[k]$ to each item $i \in [n]$ with probability $1/k$, %$f_{c_{i}}$, and then
weighting each edge $(ij)\in E$ with the measurement $s_{ij}$,
drawn from the probability density $p_{c_{i},c_{j}}$. The aim is to
recover the cluster assignments $c_{i}$ for $i\in [n]$ from the
measurement graph thus constructed.

We consider the sparse regime $\alpha = O(1)$, and the limit $n\to \infty$ with fixed number of clusters $k$. With high probability, the graph $G$ is disconnected, so that \emph{exact recovery} of the clusters, as considered e.g. in \cite{jog2015information,7282873}, is impossible. In this paper, we address instead the question of how many measurements are needed to \emph{partially recover} the cluster assignments, i.e. to infer cluster assignments $\hat{c}_{i}$ such that the following quantity, called \emph{overlap}, is strictly positive:
\begin{equation}
\label{overlap}
\max_{\sigma\in\mathfrak{S}_k}\frac{\frac{1}{n}\sum_{i} \ind(\sigma(\hat{c}_{i}) = c_{i}) - \frac{1}{k} }{1 - \frac{1}{k}}\, ,
\end{equation}
where $\mathfrak{S}_{k}$ is the set of permutations of $[k]$. This quantity is monotonously increasing with the number of correctly classified items. In the limit $n\to \infty$, it vanishes for a random guess, and equals unity if the recovery is perfect. Finally, we note an important special case for which analytical results can be derived, which is the case of symmetric clusters: $\forall a,b\in[k]$
\begin{align}
\begin{split}
%f_{a} &= 1/k\, , \\ 
 p_{a,b}(s) &:= \ind(a=b)p_{\rm in}(s) + \ind(a\neq b)p_{\rm out}(s),
\end{split}
\label{symmetric_model}
\end{align}
where $p_{\rm in}(s)$ (resp. $p_{\rm out}(s)$) is the probability density of observing a measurement $s$ between items of the same cluster (resp. different clusters). 
For this particular case, we conjecture that all of the three algorithms we propose achieve partial recovery of the clusters whenever $\alpha > \alpha_c$, where
\begin{equation}
\label{transition}
 \frac{1}{\alpha_c} = \frac{1}{k}\int_{\mathcal{K}} \d s\ \frac{\big(p_{\rm in}(s) - p_{\rm out}(s)\big)^{2}}{p_{\rm in}(s) + (k-1)p_{\rm out}(s)}\, ,
\end{equation}
where $\mathcal{K}$ is the support of the function $p_{\rm in} + (k-1)p_{\rm out}$. This expression corresponds to the threshold of a related reconstruction problem on trees \cite{mezard2006reconstruction}. 
In the following, we substantiate this claim for the case of $k\!=\!2$ symmetric clusters, and discrete measurement distributions. 
Note that the model we introduce is a special case of the labeled stochastic block model of \cite{HLM2012}. In particular, for the case $k\!=\!2$, %of two symmetric clusters,
it was proven in \cite{LMX2013} that partial recovery is information theoretically impossible if $\alpha < \alpha_c$. In this contribution, we argue that this bound is tight, namely that partial recovery is possible whenever $\alpha > \alpha_c$, and that the algorithms we propose are optimal, in that they achieve this threshold. 
% Note also that when the measurements $s_{ij}=\pm 1$ are binary variables, the symmetric model (\ref{symmetric_model}) reduces to the censored block model of \cite{abbe2014decoding}. 
Note also that the symmetric model (\ref{symmetric_model}) contains the censored block model of \cite{abbe2014decoding}.
More precisely, if $p_{\rm in}$ and $p_{\rm out}$ are discrete distributions on $\{\pm 1\}$ with $p_{\rm in}(+1)=p_{\rm out}(-1)=1-\epsilon$, then $\alpha_c=(1-2\epsilon)^{-2}$. In this case, the claimed result is known \cite{saade2015spectral}, and to the best of our knowledge, this is the only case where our result is known.

\subsection{Motivation and related work}
% There are various interpretations and models that connect to this
% problem such as i) Community detection \cite{abbe2013conditional}: we
% try to recover the community membership of the nodes based on noisy
% (or censored) observations about their relationship; ii) Correlation
% clustering \cite{bansal2004correlation}: we try to cluster the graph
% $G$ by minimizing the number of ``disagreeing edges'' ($J_{ij}=-1$) in
% each cluster. These examples, and others such as synchronisation, are
% discussed in details in \cite{abbe2014decoding}.

The ability to cluster data from as few pairwise comparisons as
possible is of broad practical interest \cite{7282873}. First, there are situations where all the pairwise comparisons are simply not available. This is particularly the case if a comparison is the result of a human-based experiment. For instance, in crowdclustering \cite{gomes2011crowdclustering,yi2012crowdclustering}, people are asked to compare a subset of the items in a dataset, and the aim is to cluster the whole dataset based on these comparisons. Clearly, for a large dataset of size $n$, we can't expect to have all $O(n^{2})$ measurements. Second, even if these comparisons can be automated, the typical cost of computing all pairwise measurements is $O(n^{2}d)$ where $d$ is the dimension of the data. For large datasets with $n$ in the millions or billions, or large dimensional data, like high resolution images, this cost is often prohibitive. Storing all $O(n^{2})$ measurements is also problematic. Our work supports the idea that if the measurements between different classes of items are sufficiently different, a random subsampling of $O(n)$ measurements might be enough to accurately cluster the data. 

This work is inspired by recent progress in the problem of detecting
communities in the sparse stochastic block model (SBM) where partial
recovery is possible only when the average degree $\alpha$ is larger
than a threshold value, first conjectured in
\cite{decelle2011asymptotic}, and proved in
\cite{mossel2012stochastic,massoulie2013community,mossel2013proof}. A
belief propagation (BP) algorithm similar to the one presented here is
introduced in \cite{decelle2011asymptotic}, and argued to be optimal
in the SBM. Spectral algorithms that match the performance of BP were
later introduced in \cite{krzakala2013spectral,saade2014spectral}. The
spectral algorithms presented here are based on a generalization of
the operators that they introduce.

\subsection{Outline and main results}

In Sec. \ref{sec:algorithms}, we describe three closely related algorithms to solve the partial recovery problem of Sec. \ref{sec:problem_and_model}. The first one is a belief propagation (BP) algorithm approximating the Bayes optimal solution. The other two are spectral methods derived from BP. We show numerically that all three methods achieve the threshold (\ref{transition}). Next in Sec. \ref{sec:spectral} we substantiate this claim for the spectral method based on the non-backtracking operator.

\begin{figure}[!t]
\begin{center}
\includegraphics[width=0.48\textwidth]{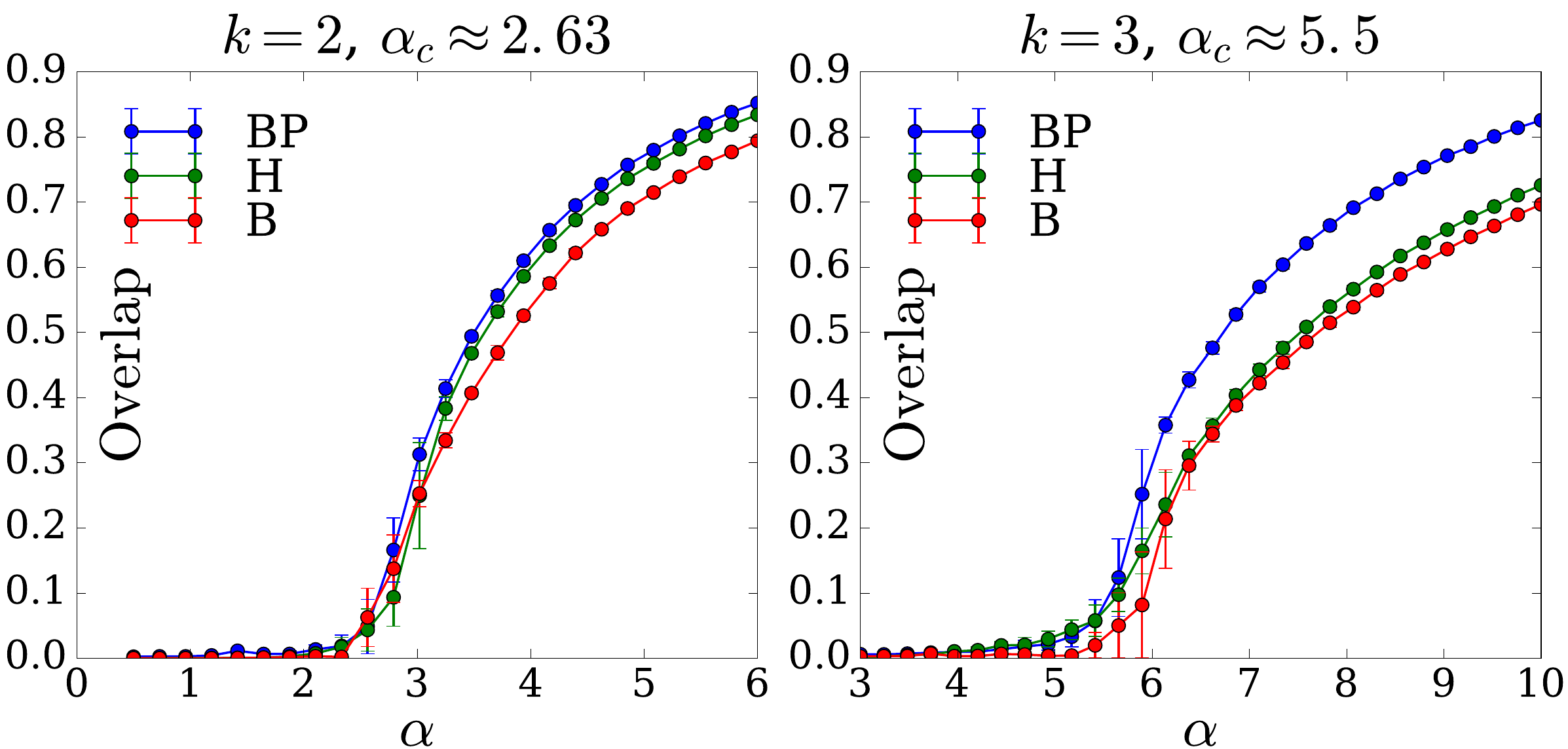}
\end{center}
\vspace{-1.2em}
\caption{Performance in clustering model-generated measurement graphs in the symmetric case (\ref{symmetric_model}). The overlap is averaged over $20$ realizations  of graphs of size $n=10^{5}$, with $k=2,3$ clusters, and Gaussian $p_{\rm in},p_{\rm out}$ with mean respectively $1.5$ and $0$, and unit variance. The theoretical transition $(\ref{transition})$ is at $\alpha_{c}\approx 2.63$ for $k=2$, and conjectured to be at $\alpha_{c}\approx 5.5$ for $k=3$. All three methods achieve the theoretical transition, although the Bethe Hessian (H) and belief propagation (BP) achieve a higher overlap than the non-backtracking operator (B).}
\label{fig:overlapVSalpha}
\end{figure}

\section{Algorithms}
\label{sec:algorithms}

\subsection{Belief propagation}

We consider a measurement graph generated from the model of Section \ref{sec:problem_and_model}. From Bayes' rule, we have:
\begin{align}
\P( \{c_{i}\} | \{s_{ij}\} ) &= \frac{1}{Z} \underset{(ij)\in{E}}{\prod} p_{c_{i},c_{j}}(s_{ij})\, ,
% \prod_{i=1}^{n} f_{c_{i}}
\end{align}
where $Z$ is a normalization. The Bayes optimal assignment, maximizing the overlap (\ref{overlap}), is $\hat{c}_{i} = \argmax\ \P_{i}$, the mode of the marginal of node $i$. We approximate this marginal using belief propagation (BP):
\begin{equation}
\label{BPmarg}
\P_{i}(c_{i}) \approx \frac{1}{Z_{i}}  \prod_{l\in \partial i} \sum_{c_{l}=1}^{k} p_{c_{i},c_{l}}(s_{il}) \P_{l\to i}(c_{l})\, , 
\end{equation}
where $\partial i$ denotes the neighbors of node $i$ in the measurement graph $G$, $Z_{i}$ is a normalization, and the $\P_{i\to j}(c_{i})$ are the fixed point of the recursion:
\begin{equation}
\label{BPrec}
\P_{i\to j}(c_{i}) = \frac{1}{Z_{i\to j}}  \prod_{l\in \partial i\backslash j} \sum_{c_{l}=1}^{k} p_{c_{i},c_{l}}(s_{il}) \P_{l\to i}(c_{l})\, .
\end{equation}
In practice, starting from a random initial condition, we iterate
(\ref{BPrec}) until convergence, and estimate the marginals from
(\ref{BPmarg}). On sparse tree-like random graphs generated by our
model, BP is widely believed to give asymptotically accurate results,
though a rigorous proof is still lacking.  This algorithm is general
and applies to any model parameters $p_{ab}$. For now on, however, we
restrict our theoretical discussion to the symmetric model
(\ref{symmetric_model}).  Eq. (\ref{BPrec}) can be written in the
compact form $\P = F(\P)$, where $\P\in\mathbb{R}^{2mk}$ and $m = |E|$
is the number of edges in $G$.

The first step in understanding the behavior of BP is to note that in
the case of symmetric clusters (\ref{symmetric_model}), there exists a
trivial fixed point of the recursion (\ref{BPrec}), namely
$\P_{i\to j}(c_{i}) = 1/k$.  This fixed point is uninformative, yielding a vanishing overlap.
% in the
% sense that it corresponds to a random guess, with vanishing
% overlap. 
If this fixed point is stable, then starting from an initial
condition close to it will cause BP to fail to recover the
clusters. We therefore investigate the linearization of (\ref{BPrec})
around this fixed point, given by the Jacobian $J_F$.

\subsection{The non-backtracking operator}
\label{sec:non_bactracking}
A simple computation yields 
\begin{equation}
J_{F} = \B\otimes\left(\I_{k} - \frac{1}{k}U_{k}\right)\, ,
\end{equation}
where $I_{k}$ is the $k\times k$ identity matrix, $U_{k}$ is the $k\times k$ matrix with all its entries equal to $1$, $\otimes$ denotes the tensor product, and $\B$ is a $2m\times 2m$ matrix called the non-backtracking operator, acting on the directed edges of the graph $G$, with elements for $(ab)$ and $(cd)\in E$:
\begin{align}
\begin{split}
  \label{B}
\B_{(a\to b),(c\to d)} &= w(s_{cd})\ind(a=d)\ind(b \neq c)\, , \\
\forall s, w(s) &:= \frac{p_{\rm in}(s) - p_{\rm out}(s)}{p_{\rm in}(s) + (k-1)p_{\rm out}(s)}\, .
\end{split}
\end{align}
 Note that to be consistent with the analysis of BP, 
our definition of the non-backtracking operator is the transpose of the definition of \cite{krzakala2013spectral}.
 % we defined the non-backtracking operator in an unconventional manner. We recover the standard non-backtracking operator by taking the transpose of the current definition.
 This matrix generalizes the non-backtracking operators of \cite{krzakala2013spectral,saade2015spectral} to arbitrary edge weights.
More precisely, for the censored block model \cite{abbe2014decoding}, we have $s=\pm 1$ and $w(s)=(1-2\epsilon)s$ so that $\B$ is simply a scaled version of the matrix introduced in \cite{saade2015spectral}. We also introduce an operator $\C\in\mathbb{R}^{n\times 2m}$ defined as 
\begin{equation}
\label{defC}\C_{i,j\to l} = w(s_{jl})\ind(i = l)\, .
\end{equation}
This operator follows from the linearization of eq. (\ref{BPmarg}) for small $\P_{l\to i}$. 
Based on these operators, we propose the following spectral algorithm. First, compute the real eigenvalues of $\B$ with modulus greater than $1$. Let $r$ be their number, and denote by $v_{1},...,v_{r}\in\mathbb{R}^{2m}$ the corresponding eigenvectors. If $r=0$, raise an error. Otherwise, form the matrix $Y = [v_{1} \cdots v_{r}]\in\mathbb{R}^{2m\times r}$ by stacking the eigenvectors in columns, and let $X = \C Y\in\mathbb{R}^{n\times r}$. Finally, regarding each item $i$ as a vector in $\mathbb{R}^{r}$ specified by the $i$-th line of $X$, cluster the items, using e.g. the k-means algorithm. 

Theoretical guarantees for the case of $k = 2$ clusters are sketched in the next section, stating that this simple algorithm succeeds in partially recovering the true clusters all the way down to the transition (\ref{transition}). Intuitively, this algorithm can be thought of as a spectral relaxation of belief propagation. Indeed, for the particular case of $k=2$ symmetric clusters, we will argue that the spectral radius of $\B$ is larger than $1$ if and only if $\alpha>\alpha_{c}$. As a simple consequence, whenever $\alpha < \alpha_{c}$, 
% the spectral radius of $J_{F}$ is also smaller than $1$, so that 
the trivial fixed point of BP is stable, and BP fails to recover the clusters. On the other hand, when $\alpha > \alpha_c$, a small perturbation of the trivial fixed point grows when iterating BP. Our spectral algorithm approximates the evolution of this perturbation by replacing the non-linear operator $F$ by its Jacobian $J_{F}$. In practice, as shown on figure \ref{fig:overlapVSalpha}, the non-linearity of the function $F$ allows BP to achieve a better overlap than the spectral method based on $\B$, but a rigorous proof that BP is asymptotically optimal is still lacking.   

\subsection{The Bethe Hessian}
\label{sec:bethe_hessian}

The non-backtracking operator $\B$ of the last section is a large, non-symmetric matrix, making the implementation of the previous algorithm numerically challenging. A much smaller, closely related symmetric matrix can be defined that empirically performs as well in recovering the clusters, and in fact slightly better than $\B$. For a real parameter $x\geq 1$, define a matrix $\H(x)\in\mathbb{R}^{n \times n}$ with non-zero elements: 
\begin{equation}
\label{Bethe_hessian}
 \H_{ij}(x) = 
 \begin{cases}
 1 + \sum_{l\in\partial i} \frac{w(s_{il})^{2}}{x^{2}  -   w(s_{il})^{2} } \text{ if } i=j  \vspace{0.5em}  \\ 
  -  \frac{x w(s_{ij})}{x^{2}  -   w(s_{ij})^{2} } \text{ if } (ij)\in E \\
 \end{cases} 
 \, ,
 % \ind(i=j)\left(1 + \sum_{l\in\partial i} \frac{w(s_{il})^{2}}{X^{2}  -   w(s_{il})^{2} }    \right)  -  \frac{X w(s_{ij})}{X^{2}  -   w(s_{ij})^{2} } \, ,
 \end{equation} 
where $\partial i$ denotes the set of neighbors of node $i$ in the graph $G$, and $w$ is defined in (\ref{B}). A simple computation, analogous to \cite{saade2015spectral}, allows to show that $(\lambda\geq 1,v)$ is an eigenpair of $\B$, if and only $\H(\lambda)v=0$. This property justifies the following picture \cite{saade2014spectral}. For $x$ large enough, $\H(x)$ is positive definite and has no negative eigenvalue. As we decrease $x$, $\H(x)$ gains a new negative eigenvalue whenever $x$ becomes smaller than an eigenvalue of $\B$. Finally, at $x = 1$, there is a one to one correspondence between the negative eigenvalues of $\H(x)$ and the real eigenvalues of $\B$ that are larger than $1$. We call Bethe Hessian the matrix $\H(1)$, and propose the following spectral algorithm, by analogy with Sec. \ref{sec:non_bactracking}. First, compute all the negative eigenvalues of $\H(1)$. Let $r$ be their number. If $r=0$, raise an error. Otherwise, denoting $v_{i},...,v_{r}\in\mathbb{R}^{n}$ the corresponding eigenvectors, form the matrix $X =[v_{1} \cdots v_{r}]\in\mathbb{R}^{n\times r}$ by stacking them in columns. Finally, regarding each item $i$ as a vector in $\mathbb{R}^{r}$ specified by the $i$-th line of $X$, cluster the items, using e.g. the k-means algorithm. 

In the case of two symmetric clusters, the results of the next section imply that if $\alpha > \alpha_c$, denoting by $\lambda_1 > 1$ the largest eigenvalue of $\B$, the smallest eigenvalue of $\H(\lambda_1)$ is $0$, and the corresponding eigenvector allows partial recovery of the clusters. While the present algorithm replaces the matrix $\H(\lambda_1)$ by the matrix $\H(1)$ and is therefore beyond the scope of this theoretical guarantee, we find empirically that the eigenvectors with negative eigenvalues of $\H(1)$ are also positively correlated with the hidden clusters, and in fact allow better recovery (see figure \ref{fig:overlapVSalpha}), without the need to build the non-backtracking operator $\B$ and to compute its leading eigenvalue. 

This last algorithm also has an intuitive justification. 
It is well known \cite{yedidia2001bethe} that BP tries to optimize the so-called Bethe free energy. In the same way $\B$ can be seen as a spectral relaxation of BP, $\H(1)$ can be seen as a spectral relaxation of the direct optimization of the Bethe free energy. In fact, it corresponds to the Hessian of the Bethe free energy around a trivial stationary point (see e.g. \cite{saade2015matrix,saade2014spectral}).

\subsection{Numerical results}

\label{sec:numerical}

Figure \ref{fig:overlapVSalpha} shows the performance of all three algorithms on model-generated problems. We consider the symmetric problem defined by (\ref{symmetric_model}) with $k=2,3$, fixed $p_{\rm in}$ and $p_{\rm out}$, chosen to be Gaussian with a strong overlap, and we vary $\alpha$. All three algorithms achieve the theoretical threshold.    

\begin{figure}[!t]
\includegraphics[width = 0.48\textwidth]{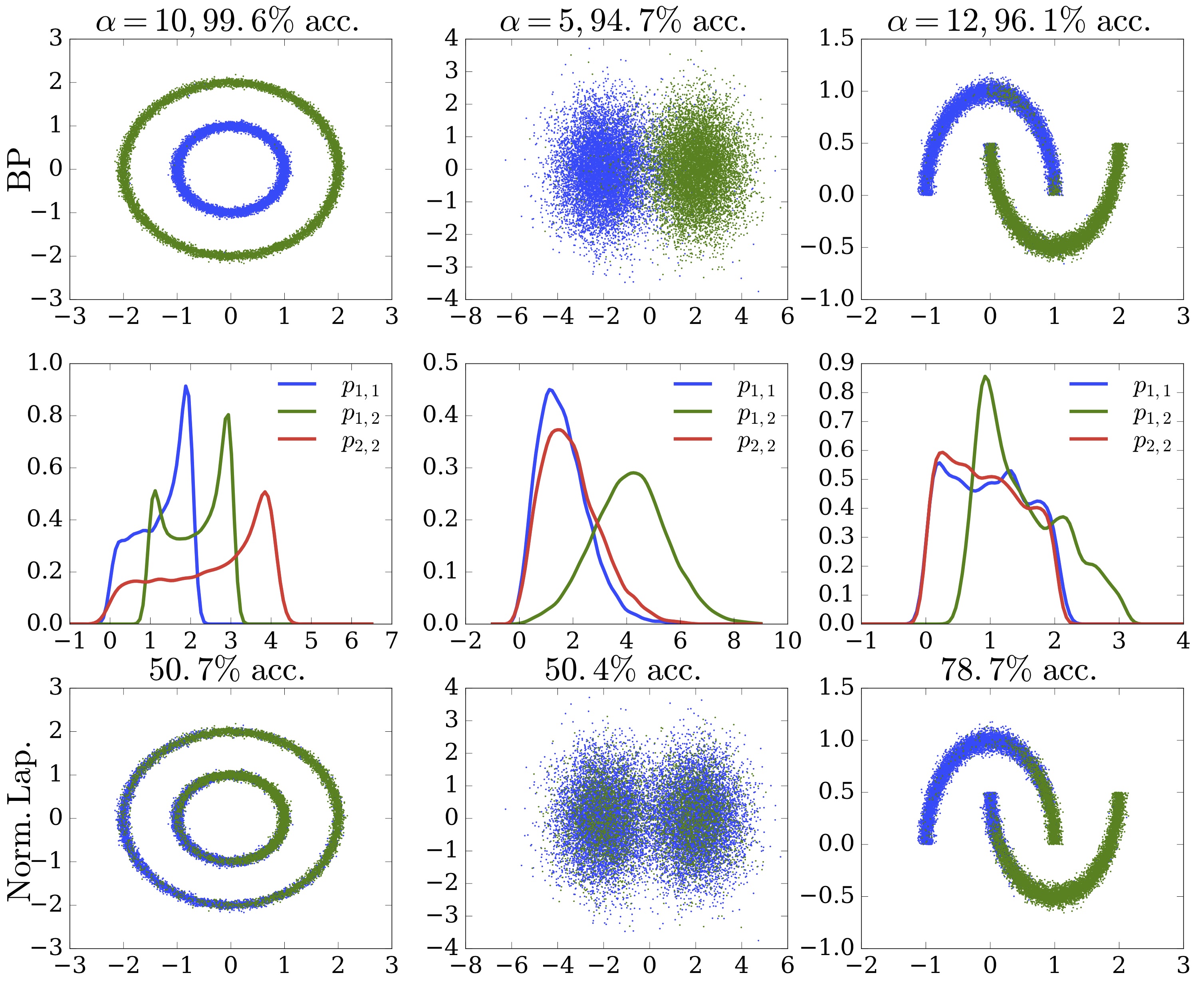}
\caption{Clustering of toy datasets using belief propagation. Each dataset is composed of $20000$ points, $200$ of which come labeled and constitute the training set. We used the Euclidean distance as the measurement function $s$, and estimated the probability densities $p_{ab}$ on the training set using kernel density estimation (middle row). Although these estimates are very noisy and overlapping, belief propagation is able to achieve a very high accuracy using a random measurement graph $G$ of small average degree $\alpha$ (top row). For comparison, we show in the third row the result of spectral clustering with the normalized Laplacian, using a $3$-nearest neighbors similarity graph (see e.g. \cite{Tutorial}) built from $G$, i.e. using only the available measurements.}
\label{fig:toy_datasets}
\end{figure}

While all the algorithms presented in this work assume the knowledge of the parameters of the model, namely the functions $p_{a,b}$ for $a,b\in[k]$, we argue that the belief propagation algorithm is robust to large imprecisions on the estimation of these parameters. To support this claim, we show on figure \ref{fig:toy_datasets} the result of the belief propagation algorithm on standard toy datasets where the parameters were estimated on a small fraction of labeled data.  

\section{Properties of the non-backtracking operator}
\label{sec:spectral}
\newcommand{\NN}{{\mathbb{N}}}

We now state our claims concerning the spectrum of ${\B}$. We restrict ourselves to the case where $k=2$ and $p_{\rm in}$ and $p_{\rm out}$ are distributions on a finite alphabet.
% although we believe our results extend to general $k$ for the symmetric model defined by (\ref{symmetric_model}).

\begin{claim}\label{the:main}
  Consider an Erd\Ho{o}s-R\'enyi random graph on $n$ vertices with average degree $\alpha$, with variables assigned to vertices $c_i\in \{1,2\}$ uniformly at random independently from the graph and measurements $s_{ij}$ between any two neighboring vertices drawn according to the probability density: $p_{c_i, c_j}(s) = \ind(c_i=c_j)p_{\rm in}(s) + \ind(c_i\neq c_j)p_{\rm out}(s)$ for two fixed (i.e. independent of $n$) discrete distributions $p_{\rm in}\neq p_{\rm out}$ on $\mathcal{S}$.
  Let $\B$ be the matrix defined by (\ref{B}) and denote by $|\lambda_1|\geq |\lambda_2|\geq \dots \geq |\lambda_{2m}|$ the eigenvalues of $\B$ in order of decreasing magnitude, where $m$ is the number of edges in the graph.
  Recall that $\alpha_c$ is defined by (\ref{transition}).
%  \begin{eqnarray*}
%\frac{1}{\alpha_c} = \frac{1}{2}\sum_{s\in \mathcal{S}} \frac{(p_{\rm in}(s)-p_{\rm out}(s))^2}{p_{\rm in}(s)+p_{\rm out}(s)}
%  \end{eqnarray*}
  Then, with probability tending to $1$ as $n\to \infty$:
  \begin{itemize}
  \item[(i)] If $\alpha<\alpha_c$, then $|\lambda_1|\leq \sqrt{\frac{\alpha}{\alpha_c}}+o(1)$.
  \item[(ii)] If $\alpha>\alpha_c$, then $\lambda_1=\frac{\alpha}{\alpha_c}+o(1)$ and $|\lambda_2|\leq \sqrt{\frac{\alpha}{\alpha_c}}+o(1)$. Additionally, denoting by $v$ the eigenvector associated with $\lambda_1$, $\C v$ is positively correlated with the planted variables $(c_i)_{i\in [n]}$, where $\C$ is defined in (\ref{defC}). 
%\begin{align}
%\BEAS
%\hat{x}_i=\frac{\text{sign}\left(\underset{j\in\partial i}{\sum} v_{j\rightarrow i}\right)+1}{2}.
%\EEAS
\end{itemize}
\end{claim}

Note that for the censored block model, our claim implies Theorem 1 in \cite{saade2015spectral}. The main idea which substantiates our claim is to introduce a new non-backtracking operator with spectral properties close to those of $\B$ and then apply the techniques developed in \cite{blm15} to it. We try to use notations consistent with \cite{blm15}:
%$\vec E$ is the set of oriented edges and for any
for an oriented edge $e=u\to v=(u,v)$ from node $u$ to node $v$, we set $e_1=u$, $e_2=v$ and
$e^{-1}=(v,u)$. For a matrice $M$, its transpose is denoted by
$M^*$. We also define $\sigma_i=2c_i-3$ for each $i\in [n]$.

We start by a simple transformation: if $t$ is the vector in
$\RR^{\vec E}$ defined by $t_e = \sigma_{e_1}$ and $\odot$ is the
Hadamard product, i.e. $(t\odot x)_e = \sigma_{e_1}x_e$, then we have
\BEA
\label{eq:trans}\B^* x = \lambda x \Leftrightarrow \B^X(t\odot x)= \lambda (t\odot x)\, ,
\EEA with $\B^X$ defined by
$\B^X_{ef}= \B_{fe}\sigma_{e_1}\sigma_{e_2}$. In particular, $\B^X$
and $\B$ have the same spectrum and there is a trivial relation
between their eigenvectors. With $X_e=\sigma_{e_1}w(s_e)\sigma_{e_2}$, we have:
\BEAS
\B^X_{ef} = X_e\ind(e_2=f_1)\ind(e_1\neq f_2)\, .
\EEAS
Moreover, note that the random variables $(A_f=\sigma_{f_1}\sigma_{f_2})_{f\in E}$ are random signs with $\PP(A_f=1)=1/2$ and the random variables $(X_f)_{f\in E}$ are such that
\BEAS
\EE X_f = \EE X_f^2 = \frac{1}{2}\sum_{s}\frac{(p_{\rm in}(s)-p_{\rm out}(s))^2}{p_{\rm in}(s)+p_{\rm out}(s)}=\frac{1}{\alpha_c}\, .
\EEAS
We now define another non-backtracking operator $\B^Y$. First, letting $\epsilon(s) = \frac{p_{\rm out}(s)}{p_{\rm in }(s)+p_{\rm out}(s)}\in [0,1]$, we define the sequence of independent random variables $\{\tilde{Y}_{e}\}_{e\in E}$ with $\PP(\tilde{Y}_e=+1|s_e)=1-\PP(\tilde{Y}_e=-1|s_e)=1-\epsilon(s_e)$, so that $\EE[\tilde{Y}_e|s_e] = w(s_e)$. We define $Y_e=\tilde{Y}_e\sigma_{e_1}\sigma_{e_2}$ and finally
\BEAS
\B^Y_{ef}=Y_f\ind(e_2=f_1)\ind(e_1\neq f_2)\, , 
\EEAS
so that $\EE[\B^Y|G,\{s_e\}_{e\in E}]=B^X$.
 % and the sequence $\{Y_e\}_{e\in E}$ is now a sequence of i.i.d. random signs with mean $\frac{1}{\alpha_c}$.
 It turns out that the analysis
 of the matrix $\B^Y$ can be done with the techniques developped in \cite{blm15}. More precisely, we define $P$ the linear mapping on $\RR^{\vec E}$ defined by
$(Px)_e=Y_ex_{e^{-1}}$ (i.e. in matrix form $P_{ef} =
Y_e\ind(f=e^{-1})$). Note that $P^*=P$ and since $Y_e^2=1$, $P$ is an
involution so that $P$ is an orthogonal matrix. A simple computation
shows that $(B^Y)^kP = P(B^Y)^{*k}$, hence $(B^Y)^kP$ is a symmetric matrix.
This symmetry corresponds to the oriented path symmetry in
\cite{blm15} and is crucial to our analysis.
If $(\tau_{j,k}), 1\leq j\leq 2m$ are the eigenvalues of $(B^Y)^kP$ and $(x_{j,k})$ is an orthonormal basis of eigenvectors, we deduce that
\BEA
\label{sing}(B^Y)^k = \sum_{j=1}^{2m} \tau_{j,k}x_{j,k}(P x_{j,k})^*\, .
\EEA
Since $P$ is an orthogonal matrix $(Px_{j,k}), 1\leq j\leq 2m$ is also an orthonormal basis of $\RR^{\vec E}$. In particular, (\ref{sing}) gives the singular value decomposition of $(B^Y)^k$. Indeed, if $t_{j,k}=|\tau_{j,k}|$ and $y_{j,k}={\rm sign}(\tau_{j,k})Px_{j,k}$, then we get
$(B^Y)^k = \sum_{j=1}^{2m} t_{j,k}x_{j,k}y^*_{j,k}$,
which is precisely the singular value decomposition of $(B^Y)^k$. As shown in \cite{blm15}, for large $k$, the decomposition (\ref{sing}) carries structural information on the graph.

A crucial element in the proof of \cite{blm15} is the result of Kesten and Stigum \cite{ks1,ks2} and we give now its extension required here which can be seen as a version of Kesten and Stigum's work in a random environment.
We write $\NN^*=\{1,2,\dots\}$ and $U=\cup_{n\geq 0} (\NN^*)^n$ the set of finite sequences composed by $\NN^*$, where $(\NN^*)^0$ contains the null sequence $\emptyset$. For $u,v\in U$, we note $|u|=n$ for the lenght of $u$ and $uv$ for the sequence obtained by the juxtaposition of $u$ and $v$. Suppose that $\{(N_u, A_{u1},A_{u2},\dots)\}_{u\in U}$ is a sequence of i.i.d. random variables with value in $\NN\times \RR^{\NN^*}$ such that $N_u$ is a Poisson random variable with mean $\alpha$ and the $A_{ui}$ are independent i.i.d. random signs with $\PP(A_{u1}=1)=\PP(A_{u1}=-1)=\frac{1}{2}$. We then define the following random variables: first $s_u$ such that $\PP(s_u|A_u=1)=p_{\rm in }(s_u)$ and $\PP(s_u|A_u=-1)=p_{\rm out}(s_u)$, then $X_{u}= A_u w(s_u)$ and $Y_u=A_u \tilde{Y}_u$ where $\PP(\tilde{Y}_u=1|s_u)=1-\PP(Y_u=-1|s_u)=1-\epsilon(s_u)$. We assume that for all $u\in U$ and $i>N_u$, $A_{ui}=s_{ui}=X_{ui}=Y_{ui}=0$.
%For simplicity, we note $(N,s_1,s_2,\dots)$ for $(N_{\emptyset},s_{\emptyset 1},s_{\emptyset 2},\dots)$.
$N_u$ will be the number of children of node $u$ and the sequence $(s_{u1},\dots, s_{uN_u})$ the measurements on edges between $u$ and its children.
We set for $u=u_1u_2\dots u_n\in U$,
\BEAS
P^X_{\emptyset} =1,&& P^X_u=X_{u_1}X_{u_1 u_2}\dots X_{u_1\dots u_n}\, ,\\
P^Y_{\emptyset} =1,&& P^Y_u=Y_{u_1}Y_{u_1 u_2}\dots Y_{u_1\dots u_n}\, .
\EEAS
We define (with the convention $\frac{0}{0}=0$)\, ,
\BEA
M_0=1,\quad M_t= \sum_{|u|=t} \frac{P^Y_u}{\alpha^t P^X_u}\, .
\EEA
Then conditionnaly on the variables $(s_u)_{u\in U}$, $M_t$ is a martingale converging almost surely and in $L^2$ as soon as $\alpha>\alpha_c$. The fact that this martingale is bounded in $L^2$ follows from an argument given in the proof of Theorem 3 in \cite{HLM2012}.

In order to apply the technique of \cite{blm15}, we need to deal with the $\ell$-th power of the non-backtracking operators. For $\ell$ not too large, the local struture of the graph (up to depth $\ell$) can be coupled to a Poisson Galton-Watson branching process, so that the computations done for $M_\ell$ above provide a good approximation of the $\ell$-th power of the non-backtracking operator and we can use the algebraic tools about perturbation of eigenvalues and eigenvectors, see the Bauer-Fike theorem in Section 4 in \cite{blm15}.

\section{Conclusion}
We have considered the problem of clustering a dataset from as few measurements as possible. On a reasonable model, we have made a precise prediction on the number of measurements needed to cluster better than chance, and have substantiated this prediction on an interesting particular case. 
We have also introduced three efficient and optimal algorithms, based on belief propagation, to cluster model-generated data starting from this transition. 
Our results suggest that clustering can be significantly sped up by using a number of measurements \emph{linear} in the size of the dataset, instead of quadratic. 
These algorithms, however, require an estimate of the distribution of the measurements between objects depending on their cluster membership. On toy datasets, we have demonstrated the robustness of the belief propagation algorithm to large imprecisions on these estimates, paving the way for broad applications in real world, large-scale data analysis. A natural avenue for future work is the unsupervised estimation of these distributions, through e.g. an expectation-maximization approach.    
% We have considered the problem of partially recovering binary
% variables from the observation of censored edge weights, and described
% two optimal spectral algorithms for this task that can provably
% perform partial recovery as soon as it is information theoretically
% possible to do so. Remarkably, these algorithms do not require the
% knowledge of the noise parameter $\epsilon$ and perform almost as well as
% belief propagation, which is expected (but not proved) to be Bayes
% optimal for this problem. This allows to close the gap from previous
% works, both algorithmically, by providing optimal spectral algorithms,
% and theoretically, by proving that the transition (\ref{transition})
% is a necessary {\it and} sufficient condition for partial recovery.

\balance

\section*{Acknowledgment}
This work has been supported by the ERC under the European Union's FP7
Grant Agreement 307087-SPARCS and by the French Agence Nationale de la
Recherche under reference ANR-11-JS02-005-01 (GAP project).

% trigger a \newpage just before the given reference
% number - used to balance the columns on the last page
% adjust value as needed - may need to be readjusted if
% the document is modified later
%\IEEEtriggeratref{8}
% The "triggered" command can be changed if desired:
%\IEEEtriggercmd{\enlargethispage{-5in}}

% references section

% can use a bibliography generated by BibTeX as a .bbl file
% BibTeX documentation can be easily obtained at:
% http://www.ctan.org/tex-archive/biblio/bibtex/contrib/doc/
% The IEEEtran BibTeX style support page is at:
% http://www.michaelshell.org/tex/ieeetran/bibtex/
\bibliographystyle{IEEEtran}
% argument is your BibTeX string definitions and bibliography database(s)
%\bibliography{IEEEabrv,../bib/paper}
%
% <OR> manually copy in the resultant .bbl file
% set second argument of \begin to the number of references
% (used to reserve space for the reference number labels box)

%\begin{thebibliography}{1}

%\bibitem{IEEEhowto:kopka}
%H.~Kopka and P.~W. Daly, \emph{A Guide to \LaTeX}, 3rd~ed.\hskip 1em plus
%  0.5em minus 0.4em\relax Harlow, England: Addison-Wesley, 1999.

%\end{thebibliography}
\bibliography{mybib}

% that's all folks
\end{document}